\documentclass[envcountsect]{llncs}
\usepackage{fullpage}

\usepackage{balance} 
\usepackage{floatrow}

\usepackage[caption=false,font=footnotesize]{subfig}
\newfloatcommand{capbtabbox}{table}[][\FBwidth]

\usepackage{amsmath}
\usepackage{bbm}
\usepackage{stmaryrd}

\usepackage{array}
\usepackage[hyphens]{url}
\usepackage[pdftitle=Title,pdfauthor=Anonymous]{hyperref}
\usepackage{ wasysym }
\usepackage{adjustbox}
\usepackage{booktabs}
\usepackage{multirow}
\usepackage{rotating}
\usepackage{makecell}
\usepackage{hhline}
\usepackage{lscape}
\usepackage{tabu}
\usepackage{tikz}
\usepackage{tikzsymbols}
\usepackage{textcomp} 
\usetikzlibrary{shapes,backgrounds} 
\usepackage[justification=centering]{caption} 
\usepackage{colortbl}

\usepackage{blindtext}
\usepackage[utf8]{inputenc}
\usepackage[T1]{fontenc}
\usepackage{amsmath}
\usepackage{amsfonts}
\usepackage{tabularx}
\usepackage{blindtext}
\usepackage{booktabs}
\usepackage{makecell}

\usepackage{cellspace}

\usepackage{longtable}

\usepackage{diagbox}

\usepackage{listings, xcolor}

\definecolor{verylightgray}{rgb}{.97,.97,.97}
\lstdefinelanguage{Solidity}{
keywords=[1]{anonymous, assembly, assert, balance, break, call, callcode, case, catch, class, constant, continue, contract, debugger, default, delegatecall, delete, do, else, event, export, external, false, finally, for, function, gas, if, implements, import, in, indexed, instanceof, interface, internal, is, length, library, log0, log1, log2, log3, log4, memory, modifier, new, payable, pragma, private, protected, public, pure, push, require, return, returns, revert, selfdestruct, send, storage, struct, suicide, super, switch, then, this, throw, transfer, true, try, typeof, using, value, view, while, with, addmod, ecrecover, keccak256, mulmod, ripemd160, sha256, sha3}, 
	keywordstyle=[1]\color{blue}\bfseries,
	keywords=[2]{Stages,States, address, bool, byte, bytes, bytes1, bytes2, bytes3, bytes4, bytes5, bytes6, bytes7, bytes8, bytes9, bytes10, bytes11, bytes12, bytes13, bytes14, bytes15, bytes16, bytes17, bytes18, bytes19, bytes20, bytes21, bytes22, bytes23, bytes24, bytes25, bytes26, bytes27, bytes28, bytes29, bytes30, bytes31, bytes32, enum, int, int8, int16, int24, int32, int40, int48, int56, int64, int72, int80, int88, int96, int104, int112, int120, int128, int136, int144, int152, int160, int168, int176, int184, int192, int200, int208, int216, int224, int232, int240, int248, int256, mapping, string, uint, uint8, uint16, uint24, uint32, uint40, uint48, uint56, uint64, uint72, uint80, uint88, uint96, uint104, uint112, uint120, uint128, uint136, uint144, uint152, uint160, uint168, uint176, uint184, uint192, uint200, uint208, uint216, uint224, uint232, uint240, uint248, uint256, var, void, ether, finney, szabo, wei, days, hours, minutes, seconds, weeks, years},	
	keywordstyle=[2]\color{teal}\bfseries,
	keywords=[3]{block, blockhash, coinbase, difficulty, gaslimit, number, timestamp, msg, data, gas, sender, sig, value, now, tx, gasprice, origin},	
	keywordstyle=[3]\color{violet}\bfseries,
	identifierstyle=\color{black},
	sensitive=false,
	comment=[l]{//},
	morecomment=[s]{/*}{*/},
	commentstyle=\color{gray}\ttfamily,
	stringstyle=\color{red}\ttfamily,
	morestring=[b]',
	morestring=[b]"
}

\usepackage{xspace}
\newcommand{\etal}{\textit{et al.}\xspace}
\newcommand{\etc}{\textit{etc.}\xspace}
\newcommand{\ie}{\textit{i.e.,}\xspace}
\newcommand{\eg}{\textit{e.g.,}\xspace}
\newcommand{\cf}{\textit{cf.}\xspace}

\newcommand\eatpunct[1]{}

\newcolumntype{?}[1]{!{\vrule width #1}}

\newcommand{\cmmnt}[1]{\ignorespaces} 

\newcommand{\cm}{\textsf{Lissy}\xspace}

\usepackage{adjustbox}

\newcommand{\headrowfrontTwo}[1]{\multicolumn{1}{c}{\adjustbox{angle=90,lap=\width-0.9em}{#1}}}

\newcommand{\headrowsidebyside}[1]{\multicolumn{1}{c}{\adjustbox{angle=75,lap=\width-0.9em}{#1}}}
\newcommand{\headrowcleaning}[1]{\multicolumn{1}{c}{\adjustbox{angle=40,lap=\width-0.9em}{#1}}}
\newcommand{\headrowperformance}[1]{\multicolumn{1}{c}{\adjustbox{angle=30,lap=\width-0.9em}{#1}}}

\newcommand{\full}{$\CIRCLE$}
\newcommand{\prt}{$\LEFTcircle$}
\newcommand{\empt}{$\Circle$}

\begin{document}

\frontmatter
\mainmatter

\title{\Large \bf Lissy: Experimenting with on-chain order books}

\author{Mahsa Moosavi \and Jeremy Clark }
\institute{Concordia University, Montréal, Canada}

\maketitle



\begin{abstract}

Financial regulators have long-standing concerns about fully decentralized exchanges that run `on-chain' without any obvious regulatory hooks. The popularity of Uniswap, an automated market makers (AMM), made these concerns a reality. AMMs implement a lightweight dealer-based trading system, but they are unlike anything on Wall Street, require fees intrinsically, and are susceptible to front-running attacks. This leaves the following research questions we address in this paper: (1) are conventional (\ie order books), secure (\ie resistant to front-running and price manipulation) and fully decentralized exchanges feasible on a public blockchain like Ethereum, (2) what is the performance profile, and (3) how much do Layer 2 techniques (\eg Arbitrum) increase performance? To answer these questions, we implement, benchmark, and experiment with an Ethereum-based call market exchange called \cm. We confirm the functionality is too heavy for Ethereum today (you cannot expect to exceed a few hundred trade executions per block) but show it scales dramatically (99.88\% gas cost reduction) on Arbitrum. 

\end{abstract}




\section{Introductory Remarks}

There are three main approaches to arranging a trade~\cite{Har03}. In a \emph{quote-driven} market, a dealer uses its own inventory to offer a price for buying or selling an asset. In a \emph{brokered exchange}, a broker finds a buyer and seller. In an \emph{order-driven} market, offers to buy (\emph{bids}) and sell (\emph{offers}/\emph{asks}) from many traders are placed as orders in an order book. Order-driven markets can be \emph{continuous}, with buyers/sellers at any time adding orders to the order book (\emph{makers}) or executing against an existing order (\emph{takers}); or they can be \emph{called}, where all traders submit orders within a window of time and orders are matched in a batch (like an auction). 

Conventional financial markets (\eg NYSE, NASDAQ) use both continuous time trading during open hours, and a call market before and during open hours to establish an opening price and a closing price. After early experiments at implementing continuous time trading on Ethereum (\eg EtherDelta, OasisDEX), it was generally accepted that conventional trading is infeasible on Ethereum for performance reasons. Centralized exchanges continued their predominance, while slowly some exchanges moved partial functionality on-chain (\eg custody of assets) while executing trades off-chain. 

A clever quote-driven alternative, called an automatic market maker (AMM), was developed that only requires data structures and traversals with low gas complexity. This approach has undesirable price dynamics (\eg market impact of a trade, slippage between the best bid/ask and actual average execution price, \etc) which explains why there is no Wall Street equivalent, however, it is efficient on Ethereum and works `good enough' to attract trading. First generation AMMs provide makers (called liquidity providers) with no ability to act on price information---they are uninformed traders that can only lose (called impermanent loss) on trades but make money on fees. Current generation AMMs (\eg Uniswap v3) provided informed makers with a limited ability (called concentrated liquidity) to act on proprietary information~\cite{Park21} without breaking Ethereum's performance limitations. Ironically, the logical extension of this is a move back to where it all started---a full-fledged order-driven exchange that allows informed makers the fullest ability to trade strategically.

\textbf{Contributions.} In this paper, we experiment with on-chain markets to understand in detail if they remain infeasible on Ethereum and what the limiting factors are. Some highlights from our research include answering the following questions:
\begin{itemize}
\item What type of exchange has the fairest price execution on balance? (A call market.)
\item How many orders can be processed on-chain? (Upper-bounded by 152 per block.)
\item How much efficiency can be squeezed from diligently choosing the best data structures? (Somewhat limited; turn 38 trades into 152.)
\item To what extent can we mitigate front-running attacks? (Almost entirely.)
\item Can we stop the exchange's storage footprint on Ethereum from bloating? (Yes, but it is so expensive that it is not worth it.)
\item Are on-chain order books feasible on layer 2? (Yes! Optimistic roll-ups reduce gas costs by 99.88\%.)
\item Which aspects of Ethereum were encountered that required deeper than surface-level knowledge to navigate? (Optimizing gas refunds, Solidity is not truly object-oriented, miner extractable value (MEV) can be leveraged for good, and bridging assets for layer 2.)
\item How hard is an on-chain exchange to regulate? (The design leaves almost no regulatory hooks beyond miners (and sequencers on layer 2).)
\end{itemize}

\section{Preliminaries}

\subsection{Ethereum}

We assume the reader is familiar with the following concepts: blockchain technology; smart contracts and decentralized applications (DApps) on Ethereum; how Ethereum transactions are structured, broadcast, and finalized; the gas model including the gas limit (approximately 11M gwei at the time of our experiments) per block. A \textbf{gas refund} is a more esoteric subject (not covered thoroughly in any academic work to our knowledge) that we use heavily in our optimizations. Briefly, certain EVM operations (\texttt{SELFDESTRUCT} and \texttt{SSTORE 0}) cost negative gas, with the follow caveats: the refund is capped at 50\% of the total gas cost of the transaction, and (2) the block gas limit applies to the pre-refunded amount (\ie a transaction receiving a full refund can cost up to 5.5M gas with an 11M limit). We provide full details of all of these topics in Appendix~\ref{app:ethback}. 

\subsection{Trade Execution Systems}

\begin{table}[t]
\centering
\begin{tabular}{|p{1.5cm}|p{5.5cm}|p{4cm}|p{3.5cm}|}
\hline
\textbf{Type}			 &\textbf{Description}   				& \textbf{Advantages}      				& \textbf{Disadvantages}                   \\ \hline

Centralized Exchanges (CEX)
& Order-driven exchange acts as a trusted third party (\eg Binance, Bitfinex)
& Conventional \newline
Highest performance \newline
Low fees \newline
Easy to regulate \newline
Low price slippage  \newline
Verbose trading strategies 
&  Fully trusted custodian \newline
Slow withdrawals \newline
Server downtime \newline
Uncertain fair execution 
\\ 
\hline

Partially On-chain Exchange
& Order-driven exchange acts as a semi-trusted party (\eg EtherDelta, 0x, IDEX, Loopring)
&  High performance \newline
Low fees \newline
Easy to regulate \newline
Low price slippage \newline
Verbose trading strategies \newline
Semi-custodial
& 
Slow withdrawals \newline
Server downtime \newline
Front-running attacks \newline
Uncertain fair execution
\\ 
\hline

On-Chain Dealers
& Quote-driven decentralized exchange trades from inventory with public pricing rule (\eg Uniswap v3)
&  Non-custodial \newline
Instant trading \newline
Moderate performance \newline
Fair execution 
&  Unconventional \newline
Impermanent loss \newline
High price slippage \newline
Intrinsic fees \newline
Front-running attacks \newline
Limited trading strategies \newline
Hard to regulate
\\ 
\hline

On-chain Order-Driven Exchanges 
& Order-driven decentralized exchange executes trades between buyers and sellers (\eg \cm)
&  Conventional \newline
Non-custodial \newline
Low price slippage \newline
Fair execution \newline
Verbose trading strategies \newline
Front-running is mitigable 
& Very low performance \newline
Hard to regulate
\\ 
\hline
	
\end{tabular}
\caption{ Comparison among different trade execution systems.
\label{tab:eval2}}
\end{table}


Table~\ref{tab:eval2} illustrates various trade execution systems and summarizes their advantages and disadvantages. Appendix~\ref{app:markets} provides a full justification for the table. Briefly, fully decentralized, on-chain exchanges require the lowest trust, provide instant settlement, and have transparent trading rules that will always execute correctly. Front-running attacks (see Section~\ref{sec:front} for a very thorough discussion) are weaknesses inherent in blockchains that require specific mitigation.  

\subsection{Related Work}

Call markets are studied widely in finance and provide high integrity prices (\eg closing prices that are highly referenced and used in derivative products)~\cite{HS04,PS03,FH21}. They can also combat high frequency trading~\cite{budish2015high,aquilina2020quantifying}. An older 2014 paper~\cite{clark2014decentralizing} on the `Princeton prediction market'~\cite{Bra13} show that call markets  mitigate most blockchain-based front-running attacks present in an on-chain continuous-trading exchange as well as other limitations: block intervals are slow and not continuous, there is no support for accurate time-stamping, transactions can be dropped or reordered by miners, and fast traders can react to submitted orders/cancellations when broadcast to network but not in a block and have their orders appear first. The paper does not include an implementation, was envisioned as running on a custom blockchain (Ethereum was still in development in 2014) and market operations are part of the blockchain logic. 

The most similar academic work to this paper is the Ethereum-based periodic auction by Galal \etal~\cite{galalpublicly} and the continuous-time exchange TEX~\cite{khalil2019tex}. As with us, front-running is a main consideration of these works. In a recent SoK on front-running attacks in blockchain~\cite{eskandari2019sok}, three general mitigations are proposed: confidentiality, sequencing, and design. Both of these papers use confidentiality over the content of orders~(\cf \cite{TP07,YSLT10,TW12,cartlidge2019mpc,massacci2018futuresmex}). The main downside is that honest traders cannot submit their orders and leave, they must interact in a second round to reveal their orders. The second mitigation approach is to sequence transactions according to some rule akin to first-in-first-out~\cite{kelkar2020order,Kla20}. These are not available for experimentation on Ethereum yet (although Chainlink has announced an intention\footnote{A. Juels. \href{https://blog.chain.link/chainlink-fair-sequencing-services-enabling-a-provably-fair-defi-ecosystem/}{blog.chain.link}, 11 Sep 2020.}). The third solution is to design the service in a way that front-running attacks are not profitable---this is the approach with \cm which uses \textit{no cryptography} and is \textit{submit-and-go} for traders. A detailed comparison of front-running is provided in Section~\ref{sec:front}. Our paper also emphasizes implementation details: Galal \etal do not provide a full implementation, and TEX uses both on-chain and off-chain components, and thus does not answer our research question of how feasible an on-chain order book is.


\section{Call Market Design}


\begin{table}[t]

\centering
\begin{tabular}{|>{\centering}m{2.2cm} |>{\centering\arraybackslash}m{9cm}|}
\hline
\multicolumn{1}{|c|}{\textbf{Operation}} & \multicolumn{1}{c|}{\textbf{Description}}                            			\\ \hline

	depositToken()                           & Deposits ERC20 tokens in \cm smart contract \\ \hline
	depositEther()                           	& Deposits ETH in \cm smart contract                        			 \\ \hline
	openMarket()                             & Opens the market                                                    					 \\ \hline
	closeMarket()                            & Closes the market and processes the orders       \\ \hline
	submitBid()                              & Inserts the upcoming bids inside the priority queue   \\ \hline
	submitAsk()                              & Inserts the upcoming asks inside the priority queue  \\ \hline
	claimTokens()                           & Transfers tokens to the traders                       			\\ \hline
	claimEther()                             & Transfers ETH to the traders                       			\\ \hline
\end{tabular}
\caption{Primary operations of \cm smart contract.
\label{tab:cm_functions}}
\end{table}


A call market opens for traders to submit bids and asks which are enqueued until the market closes. Trades are executed by matching the best priced bid to the best priced ask until the best bid is less than the best ask, then all remaining trades are discarded. See Appendix~\ref{app:cm} for a numeric example. If Alice's bid of \$100 is executed against Bob's ask of \$90, Alice pays \$100, Bob receives \$90 and the \$10 difference (called a price improvement) is given to miners for reasons in explained in the front-running evaluation (Section~\ref{sec:front}).

For our experiments and measurements, we implement a call market from scratch. \cm will open for a specified period of time during which it will accept a capped number of orders (\eg 100 orders---parameterized so that all orders can be processed), and these orders are added to a priority queue (discussed in Section~\ref{sec:pq}). Our vision is the market would be open for a very short period of time, close, and then reopen immediately (\eg every other block). \cm is open source and written in 336 lines (SLOC) of Solidity plus the priority queue (\eg we implement 5 variants, each around 300 SLOC). We tested it with the Mocha testing framework using Truffle~\cite{TruffleO71:online} on Ganache-CLI~\cite{GanacheT25:online} to obtain our performance metrics. Once deployed, the bytecode of \cm is 10,812 bytes plus the constructor code (6,400 bytes) which is not stored. The Solidity source code for \cm and Truffle test files are available in a GitHub repository.\footnote{https://github.com/MadibaGroup/2020-Orderbook} We have also deployed \cm on Ethereum's testnet Rinkeby with flattened (single file) source code of just the \cm base class and priority queue implementations. It is visible and can be interacted with here: \href{https://rinkeby.etherscan.io/address/0x0d91de29c531d074853a5cef7cf9dfeb9c6ec4e0}{[etherscan.io]}. We cross-checked for vulnerabilities with \textit{Slither}\footnote{https://github.com/crytic/slither} and \textit{SmartCheck}\footnote{https://tool.smartdec.net} and it only fails some `informational' warnings that are intentional design choices (\eg a costly loop). All measurements assume a block gas limit of $11\,741\,495$ and 1 gas $=$ 56 Gwei.\footnote{EthStats (July 2020): \url{https://ethstats.net/}} Table~\ref{tab:cm_functions} summarizes \cm's primary operations.\


%
%



\subsection{Priority Queues}\label{sec:pq}

In designing \cm within Ethereum's gas model, performance is the main bottleneck. For a call market, closing the market and processing all the orders are the most time-consuming steps. Assessing which data structures will perform best is hard (\eg gas refunds, a relatively cheap mapping data structure, only partial support for object-oriented programming) without actually deploying and evaluating several variants.

We first observe that orders are executed in order: highest to lowest price for bids, and lowest to highest price for asks. This means random access to the data structure holding the orders is unnecessary (we discuss cancelling orders later in Section~\ref{sec:cancel}). We can use a lightweight \textit{priority queue} (PQ) which has only two functions: \textsf{Enqueue()} inserts an element into the priority queue; and \textsf{Dequeue()} removes and returns the highest priority element. Specifically, we use two PQs---one for bids, where the highest price is the highest priority, and one for asks, where the lowest price is the highest priority. 

As closing the market is very expensive with any PQ, we rule out sorting the elements while dequeuing and sort during each enqueue. We then implement the following 5 PQ variants:


\begin{enumerate}

\item \textbf{Heap with Dynamic Array.} A heap is a binary tree where data is stored in nodes in a specific order where the root always represents the highest priority item (\ie highest bid price/lowest ask price). Our heap stores its data in a Solidity-provided dynamically sized array. The theoretical time complexity is logarithmic enqueue and logarithmic dequeue.

\item \textbf{Heap with Static Array.} This variant replaces the dynamic array with a Solidity storage array where the size is statically allocated. This is asymptotically the same and marginally faster in practice.  

\item \textbf{Heap with Mapping.} In this variant, we store a key for the order in the heap instead of the entire order itself. Once a key is dequeued, the order struct is drawn from a Solidity mapping (which stores key-value pairs very efficiently). This is asymptotically the same and faster with variable-sized data. 

\item \textbf{Linked List.} In this variant, elements are stored in a linked list (enabling us to efficiently insert a new element between two existing elements during enqueue). Solidity is described as object-oriented but the Solidity equivalent of an object is an entire smart contract. Therefore, an object-oriented linked list must either (1) create each node in the list as a struct---but this is not possible as Solidity does not support recursive structs---or (2) make every node in the list its own contract. The latter option seems wasteful and unusual, but it surprisingly ends up being the most gas efficient data structure to dequeue. The theoretical time complexity is linear enqueue and constant dequeue.

\item \textbf{Linked List with Mapping.} Finally, we try a variant of a linked list using a Solidity mapping. The value of the mapping is a struct with the incoming order's data and the key of the next (and previous) node in the list. The contract stores the key of the first node (head) and last node (tail) in the list. Asymptotically, it is linear enqueue and constant dequeue. 

\end{enumerate}

\begin{figure}[t]
\begin{floatrow}
\ffigbox{%
\includegraphics[width=6cm]{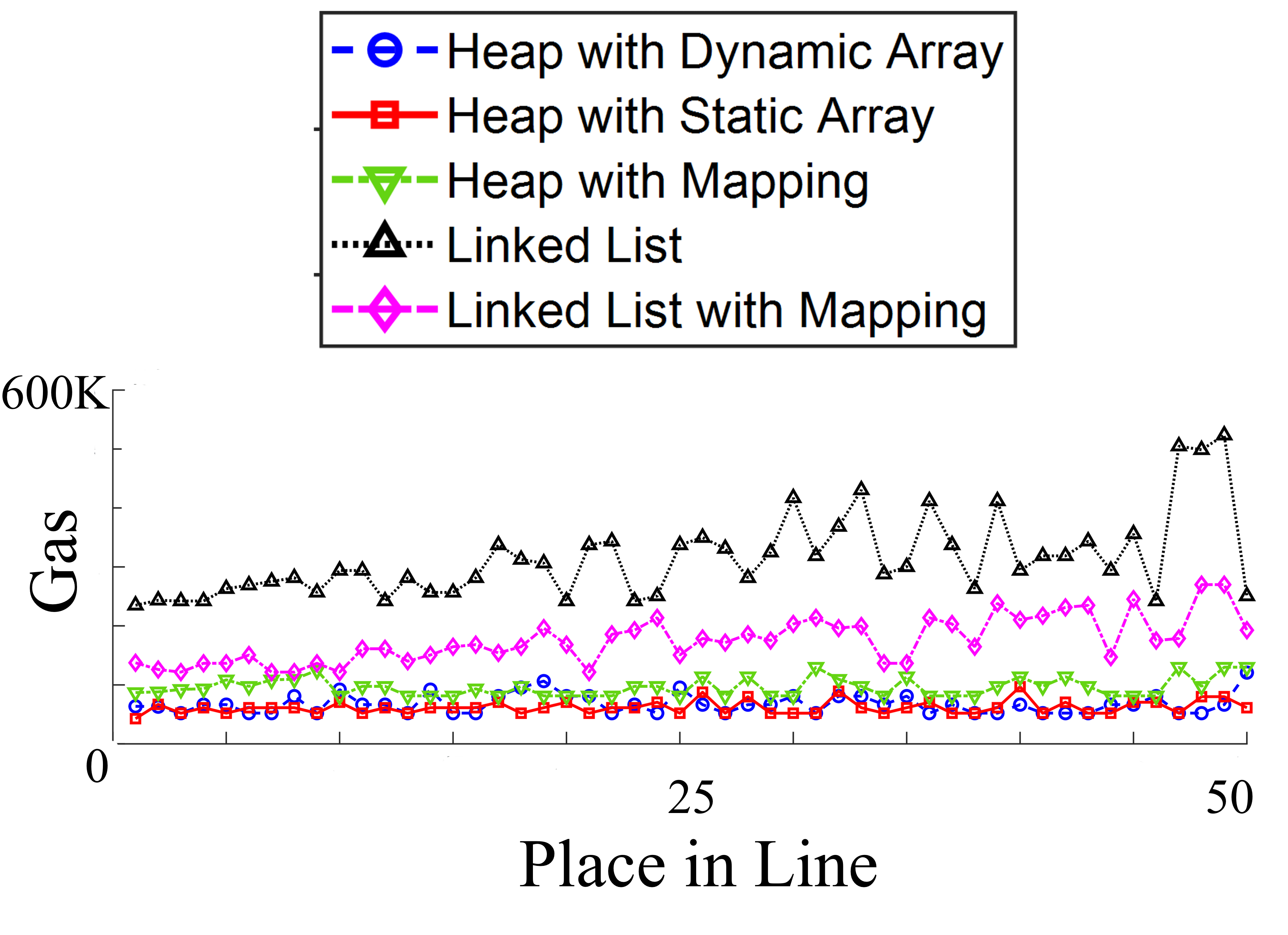}%
}{%
  \caption{\footnotesize{ Gas costs for enqueuing 50 random integers into five priority queue variants. For the x-axis, a value of 9 indicates it is the 9th integer entered in the priority queue. The y-axis is the cost of enqueuing in gas}. \label{fig:random_insertion}}%
  
}
\capbtabbox{%
\scriptsize
\begin{tabular}{|>{\centering}m{2cm} |>{\centering}m{1.15cm} |>{\centering}m{1.15cm} |>{\centering\arraybackslash}m{0.2cm}|}

\multicolumn{1}{c}{} & \headrowsidebyside{\scriptsize{\shortstack{Gas Used}}} & \headrowsidebyside{\scriptsize{Refund}} & \headrowsidebyside{\scriptsize{Full Refund?}} \\ \hline

\shortstack{Heap with\\Dynamic Array}        	& 2,518,131          & 750,000     &\full                  \\ \hline
\shortstack{Heap with\\ Static Array}          	& 1,385,307                             & 750,000      &\full                \\ \hline
\shortstack{Heap with\\ Mapping}  		& 2,781,684                            & 1,500,000    &\full                 \\ \hline
Linked List                     		& 557,085               	           & 1,200,000      &\full                \\ \hline
\shortstack{Linked List with\\ Mapping}       	& 731,514              	     	  &  3,765,000      &\full                 \\ \hline

\end{tabular}
}{%
 \caption{\footnotesize{The gas metrics associated with dequeuing 50 integers from five priority queue variants. Full refund amount is shown but the actual refund that is applied is capped.}
\label{tab:PQUnitTests}}%
}
\end{floatrow}
\end{figure}

We implemented, deployed, and tested each PQ. A simple test of enqueuing 50 integers chosen at random from a fixed interval is in Figure~\ref{fig:random_insertion} and dequeing them all is in Table~\ref{tab:PQUnitTests}.  Dequeuing removes data from the contract's storage resulting in a gas refund. Based on our manual estimates,\footnote{EVM does not expose the refund counter. We determine how many storage slots are being cleared and how many smart contracts destroyed, then we multiply these numbers by 24,000 or 15,000 respectively.} every variant receives the maximum gas refund possible (\ie half the total cost of the transaction). In other words, each of them actually consumes twice the \texttt{gasUsed} amount in gas before the refund. However, none of them are better or worse based on how much of a refund they generate.





We observe that (1) the linked list variants are materially cheaper than the heap variants at dequeuing; (2) dequeuing in a call market must be done as a batch, whereas enqueuing is paid for one at a time by the trader submitting the order; and (3) Ethereum will not permit more than hundreds of orders so asymptotic behaviour is not significant. For these reasons, we suggest using one of the linked list variants. As it can be seen in Figure~\ref{fig:random_insertion}, the associated cost for inserting elements into a linked list PQ is significantly greater than the linked list with mapping, as each insertion causes the creation of a new contract. Accordingly, we choose to implement the call market with the linked list with mapping which balances a moderate gas cost for insertion (\ie order submission) with one for removal (\ie closing the market and matching the orders). In Section~\ref{sec:rollups}, we implement \cm on Layer 2. There, the PQ variant does not change the layer 1 gas costs (as calldata size is the same) and the number of orders can be substantially increased. thus, we reconsider asymptotic and choose a heap (with dynamic array) to lower L2 gas costs across both enqueuing and dequeuing.


\subsection{Cost/Benefit of Cleaning Up After Yourself}
\label{sec:gasrefund}



\begin{table}[t]
\setlength{\tabcolsep}{0.1\tabcolsep}
\centering
\begin{tabular}{|>{\centering}m{7cm} |>{\centering}m{1.5cm} |>{\centering}m{1.5cm} |>{\centering\arraybackslash}m{0.5cm}|}

\multicolumn{1}{c}{} & \headrowcleaning{Gas Used} & \headrowcleaning{Potential Refund} & \headrowcleaning{Full Refund?} \\ \hline

Linked List without \texttt{SELFDESTRUCT}        		& 721,370          & 0     &\prt  \\ \hline
Linked List with \texttt{SELFDESTRUCT}			& 557,085          & 1,200,000     &\full  \\ \hline
Linked List with Mapping and without \texttt{DELETE}    & 334,689          & 765,000     &\full  \\ \hline
Linked List with Mapping and \texttt{DELETE}		& 731,514          & 3,765,000     &\full  \\ \hline

\end{tabular}
\caption{The gas metrics associated with dequeuing 50 integers from four linked list variants. For the refund, (\full) indicates the  refund was capped at the maximum amount and (\prt) means a greater refund would be possible.\label{tab:cleaning}}
\end{table}


%
%
%




%
%





One consequence of a linked list is that a new contract is created for every node in the list. Beyond being expensive for adding new nodes (a cost that will be bared by the trader in a call market), it also leaves a large footprint in the active Ethereum state, especially if we leave the nodes on the blockchain in perpetuity (\ie we just update the head node of the list and leave the previous head `dangling'). However in a PQ, nodes are only removed from the head of the list; thus the node contracts could be `destroyed' one by one using an extra operation, \texttt{SELFDESTRUCT}, in the \texttt{Dequeue()} function. As shown in Table~\ref{tab:cleaning}, the refund from doing this outweighs to the cost of the extra computation: gas costs are reduced from 721K to 557K.  This suggests a general principle: cleaning up after yourself will pay for itself in gas refunds. Unfortunately, this is not universally true as shown by applying the same principle to the linked list with mapping.

Dequeuing in a linked list with mapping can be implemented in two ways. The simplest approach is to process a node, update the head pointer, and leave the `removed' node's data behind in the mapping untouched (where it will never be referenced again). Alternatively, we can call \texttt{DELETE} on each mapping entry once we finish processing a trade. As it can be seen in the last two rows of Table~\ref{tab:cleaning}, leaving the data on the blockchain is cheaper than cleaning it up.

The lesson here is that gas refunds incentivize developers to clean up storage variables they will not use again, but it is highly contextual as to whether it will pay for itself. Further, the cap on the maximum refund means that refunds are not fully received for large cleanup operations (however removing the cap impacts the miners' incentives to include the transaction).  In Appendix~\ref{app:clean}, we present a second case study of the cost-benefit of clearing a mapping when it is no longer needed (including our idea to store the mapping in its own contract so it can \texttt{SELFDESTRUCT} with a single function call). The unfortunate takeaway is, again, that it is cheapest to leave the mapping in place. Cleaning up EVM state is a complicated and under-explored area of Ethereum in the research literature. For our own work, we strive to be good citizens of Ethereum and clean up to the extent that we can---thus all PQs in Table~\ref{tab:PQUnitTests} implement some cleanup.


 \subsection{\cm Performance Measurements}



\begin{table}[t]
\setlength{\tabcolsep}{0.1\tabcolsep}
\centering
\begin{tabular} {|>{\centering}m{5cm} |>{\centering}m{0.5cm} |>{\centering}m{1.5cm} |>{\centering}m{1.8cm} |>{\centering\arraybackslash}m{1.5cm}|}

\multicolumn{1}{c}{} & 

\headrowperformance{Max Trades (w.c.)} & 
\headrowperformance{Gas Used for Max Trades } & 
\headrowperformance{Gas Used for 1000 Trades } & 
\headrowperformance{Gas Used for Submission(avg)} \\ \hline

Heap with Dynamic Array       & 38            & 5,372,679                   	& 457,326,935      & 207,932     \\ \hline
Heap with Static Array         	& 42            & 5,247,636                  	& 333,656,805         & 197,710         \\ \hline
Heap with Mapping 		& 46           	 & 5,285,275                     & 226,499,722    & 215,040           \\ \hline
Linked List                     	& 152            & 5,495,265                     & 35,823,601         & 735,243             \\ \hline
Linked List with Mapping     	& 86             & 5,433,259                     & 62,774,170        &  547,466              \\ \hline

\end{tabular}
\caption{Performance of \cm for each PQ variant. Each consumes just under the block gas limit ($\sim$11M gas) with a full refund of half of its gas.\label{tab:worst_case_matching}}

\end{table}

The main research question is how many orders can be processed under the Ethereum block gas limit. The choice of PQ implementation is the main influence on performance and the results are shown in Table~\ref{tab:worst_case_matching}. These numbers are for the \textit{worst-case}---when every submitted bid and ask is marketable (\ie will require fulfillment). In practice, once \texttt{closeMarket()} hits the first bid or ask that cannot be executed, it can stop processing all remaining orders. Premised on Ethereum becoming more efficient over time, we were interested in how much gas it would cost to execute 1000 pairs of orders, which is given in the third column. The fourth column indicates the cost of submitting a bid or ask --- since this cost will vary depending on how many orders are already submitted (recall Figure~\ref{fig:random_insertion}), we average the cost of 200 order submissions.

The main takeaway is that call markets appear to be limited to processing about a hundred orders per transaction and even that is at the enormous cost of monopolizing an entire Ethereum block just to close the market. Perhaps \cm can work today in some circumstances like very low liquidity tokens, or markets with high volumes and a small number of traders (\eg liquidation auctions).


\section{\cm on Arbitrum}
\label{sec:rollups}

\textit{Layer 2 (L2)} solutions~\cite{gudgeon2020sok} are a group of scaling technologies proposed to address specific drawbacks of executing transactions on Ethereum, which is considered \textit{Layer 1 (L1)}. Among these proposals, \textit{roll-ups} prioritize reducing gas costs (as opposed to other valid concerns like latency and throughput, which are secondary for \cm). We review two variants, \textit{optimistic roll-ups} and \textit{zk roll-ups}, in Appendix~\ref{app:rollup}. Briefly, in a roll-up, every transaction is stored (but not executed) on Ethereum, then executed off-chain, and the independently verifiable result is pushed back to Ethereum, with some evidence of being executed correctly. In the Appendix, we also compare \cm on Arbitrum to Loopring 3.0.

We choose to experiment with \cm on the optimistic rollup Arbitrum.\footnote{See \url{https://offchainlabs.com} for more current details than the 2018 \textit{USENIX Security} paper~\cite{kalodner2018arbitrum}.} To deploy a DApp on Arbitrum, or to execute a function on an existing Arbitrum DApp, the transaction is sent to an \textit{inbox} on L1. It is not executed on L1, it is only recorded (as calldata) in the inbox. An open network of \textit{validators} watch the inbox for new transactions. Once inbox transactions are finalized in an Ethereum block, validators will execute the transactions and assert the result of the execution to other validators on a sidechain called \textsf{ArbOS}. As the Inbox contract maintains all Arbitrum transactions, anyone can recompute the entire current state of the ArbOS and file a dispute if executions are not correctly reported on ArbOS. Disputes are adjudicated by Ethereum itself and require a small, constant amount of gas, invariant to how expensive the transaction being disputed is. When the dispute challenge period is over, the new state of ArbOS is stored as a checkpoint on Ethereum.

\subsection{\cm Performance Measurements on Arbitrum}

\begin{table}[t]
\centering
\footnotesize

\begin{tabular} {|>{\centering}m{3cm}|>{\centering}m{3cm}|>{\centering}m{3cm}|}

 \multicolumn{1}{c}{} 								&  \multicolumn{1}{c}{\textbf{Layer1 gasUsed}}										&\multicolumn{1}{c}{\textbf{Layer2 ArbGas}} 	\tabularnewline \hline
\cm on Ethereum 				         		& 5,372,679              	   									& N/A							\tabularnewline \hline
\cm on Arbitrum 				      & 6,569 													& 508,250								\tabularnewline \hline

\end{tabular}
\caption{Gas costs of closing a market on Ethereum and on Arbitrum. ArbGas corresponds to Layer 2 \textit{computation used}.\label{tab:arbitrum_performance}}

\end{table}

\textit{Testing Platforms.} We implement \cm using the Arbitrum Rollup chain hosted on the Rinkeby testnet.  It is visible and can be interacted with here: \href{https://rinkeby-explorer.arbitrum.io/address/0x0aa5449a9f7fa34a81ce1dc720563938a27e8b03}{[Arbitrum Explorer]}. To call functions on \cm, traders can (1) send transactions directly to the \href{https://rinkeby.etherscan.io/address/0x578BAde599406A8fE3d24Fd7f7211c0911F5B29e}{Inbox contract}, or (2) use a relay server (called a \textit{Sequencer}) provided by the Arbitrum. The sequencer will group, order, and send all pending transactions together as a single Rinkeby transaction to the Inbox (and pays the gas).

In our \cm variant on Arbitrum, the validators do all computations (both enqueuing and dequeuing) so we choose to use a heap with dynamic array for our priority queue, which balances the expense of both operations. Heaps are 32\% more efficient than linked lists for submitting orders and 29\% less efficient for closing. Recall that without a roll-up, such a priority queue can only match 38 pairs at a cost of 5,372,679 gas. Table~\ref{tab:arbitrum_performance} shows that 38 pairs cost only 6,569 in L1 gas (a 99.88\% savings). This is the cost of submitting the \texttt{closeMarket()} transaction to the Inbox to be recorded, which is 103 bytes of calldata. Most importantly, recording \texttt{closeMarket()} in the Inbox will always cost around 6,569 even as the number of trades increases from 38 pairs to thousands or millions of pairs. Of course, as the number of trades increase, the work for the validators on L2 increases, as measured in ArbGas. The price of ArbGas in Gwei is not well established but is anticipated to be relatively cheap.  Arbitrum also reduces the costs for traders to submit an order: from 207,932 to 6,917 in L1 gas. In Appendix~\ref{app:rollup}, the full interaction is shown in Figure~\ref{fig:lissyl2}, which illustrates how traders interact with \cm on Arbitrum including bridges, inboxes, sequencers and validators.

Running \cm on Arbitrum has one large caveat. If the ERC20 tokens being traded are not issued on ArbOS, which is nearly always the case today, they first need to be \textit{bridged} onto ArbOS, as does the ETH. Traders send ETH or tokens to Arbitrum's bridge contracts which create the equivalent amount at the same address on L2. Withdrawals work the same way in reverse, but are only final on L1 after a dispute challenge period (currently 1 hour).\footnote{L1 users might accept assets before they are finalized as they can determine their eventual emergence on L1 is indisputable (\textit{eventual finality}).}


\section{Front-running Evaluation} 
\label{sec:front}


\begin{table*}[t]

\centering

\begin{tabular}{|c|c|c|c|c|c|c|c|c|c|c|c|c|}

\multicolumn{1}{c}{} &
\multicolumn{1}{c}{} &
\multicolumn{1}{c}{} &
\headrowfrontTwo{\shortstack{Centralized Continuous\\ Market (Coinbase)}} &
\headrowfrontTwo{\shortstack{Partially Off-chain Continuous\\ Market (EtherDelta)}} & 
\headrowfrontTwo{\shortstack{Partially Off-chain Continuous \\Market w/ Roll-up (Loopring)}} & 
\headrowfrontTwo{\shortstack{On-chain Continuous \\Market (OasisDex)}}  &
\headrowfrontTwo{\shortstack{On-chain Dark \\Continuous Market (TEX)}} &
\headrowfrontTwo{\shortstack{On-chain Automated \\Market Maker (Uniswap)}}  & 
\headrowfrontTwo{\shortstack{On-chain Call Market \\w/ Price Improvement}} &
\headrowfrontTwo{\shortstack{On-chain Call Market (\cm)}}  & 
\headrowfrontTwo{\shortstack{On-chain Call Market  \\w/ Roll-up (\cm variant)}}  &
\headrowfrontTwo{\shortstack{On-chain Dark \\Call Market (Galal \etal)}}
\\

\hline

\multicolumn{3}{|c?{1mm}}{\makecell{Who is Mallory? \\ \underline{A}uthority,  \underline{T}rader, \underline{M}iner, \underline{S}equencer	}}   	&A&A,T,M&A,T,M,S&T,M&T,M&T,M&T,M&T,M&T,M,S	&T,M	 \\ \cmidrule[1mm]{1-13}


\multirow{17}{*}{\rotatebox[origin=c]{90}{\textbf{Attack Example}}}		&\multicolumn{1}{c|}{\makecell{Mallory (\textit{maker}) squeezes in a\\ transaction before Alice's  (\textit{taker})  order}}  						&\multicolumn{1}{c?{1mm}}{Ins.}		&\empt	&\empt &\empt &\empt &\full &\empt &\full &\full  &\full &\full		\\ \hhline{~------------}					 					
								 											
&\multicolumn{1}{c|}{\makecell{Mallory (\textit{taker}) squeezes in a \\transaction before Bob's (\textit{taker 2})}   }	  								&\multicolumn{1}{c?{1mm}}{  Disp.}				&\empt &\empt &\empt &\empt &\full &\empt &\full &\full &\full	&\full		\\ \hhline{~------------}	

&\multicolumn{1}{c|}{\makecell{Mallory (\textit{maker 1}) suppresses a better \\ incoming order from Alice (\textit{maker 2}) \\ until Mallory's order is executed}}																&\multicolumn{1}{c?{1mm}}{Supp.} 				&\empt &\empt &\empt &\full &\full &\full &\prt &\prt &\prt &\prt	\\ 	\hhline{~------------}				 											

&\multicolumn{1}{c|}{\makecell{A hybrid attack based on the above \\ (\eg sandwich attacks, scalping)}}																		&\multicolumn{1}{c?{1mm}}{I/S/D}																		&\empt &\empt &\empt &\empt &\full &\empt &\empt &\full &\full	&\full												\\ 	\hhline{~------------}

&\multicolumn{1}{c|}{\makecell{Mallory suspends the market \\for a period of time}}																										&\multicolumn{1}{c?{1mm}}{Supp.}    				&\empt  &\empt  &\empt  &\prt  &\prt  &\prt  &\prt  & \prt &\prt &\prt 	\\ 	\hhline{~------------}

&\multicolumn{1}{c|}{\makecell{Spoofing: Mallory (\textit{maker}) puts an \\ order as bait, sees Alice (\textit{taker}) \\ tries to execute it, and cancels it first}}													&\multicolumn{1}{c?{1mm}}{S\&D} 				&\empt &\empt &\empt &\empt &\full &\empt &\full &\full &\full &\full \\ 	\hhline{~------------}

&\multicolumn{1}{c|}{\makecell{Cancellation Griefing: Alice (\textit{maker}) \\ cancels an order and Mallory \\ (\textit{taker}) fulfills it first}}													&\multicolumn{1}{c?{1mm}}{Disp.} 							&\empt &\empt &\empt &\empt &\full &\empt &\full & \full &\full	&\full \\ 	\hline

\end{tabular}


\caption{An evaluation of front-running attacks (rows) for different types of order books (columns). Front-running attacks are in three categories: \underline{Ins}ertion, \underline{disp}lacement, and \underline{supp}ression. A full dot (\full) means the front-running attack is mitigated or not applicable to the order book type, a partial mitigation (\prt) is awarded when the front-running attack is possible but expensive, and we give no award (\empt) if the attack is feasible. 
\label{tab:front}}
\end{table*}

As we illustrate in Table~\ref{tab:front}, call markets have a unique profile of resilience against \emph{front-running attacks}~\cite{clark2014decentralizing,eskandari2019sok,daian2019flash} that differs somewhat from continuous-time markets and automated market makers. Traders are sometimes distinguished as \emph{makers} (adds orders to a market) and \emph{takers} (trades against a pre-existing, unexecuted orders). A continuous market has both. All traders using an automated market maker are takers, while the investors who provide tokens to the AMM (liquidity providers) are makers. Under our definition, a call market only has makers: the only way to have a trade executed is to submit an order. The front-running attacks in Table~\ref{tab:front} are subcategorized, using a recent SoK~\cite{eskandari2019sok}, as being \emph{Insertion}, \emph{Displacement}, and \emph{Suppression}. To explain the difference, we will illustrate the first three attacks in the table.

In an \emph{insertion attack}, Mallory learns of a transaction from Alice. Consider Alice submitting a bid order for 100 tokens at any price (market order). Mallory decides to add new ask orders to the book (limit orders) at the maximum price reachable by Alice's order given the rest of the asks in the book. Mallory must arrange for her orders to be added before Alice's transaction and then arrange for Alice's transaction to be the next (relevant) transaction to run (\eg before competing asks from other traders are added).

In a centralized exchange, Mallory would collude with the \emph{authority} running the exchange to conduct this attack. On-chain, Mallory could be a fast \emph{trader} who sees Alice's transaction in the mempool and adds her transaction with a higher gas fee to bribe miners to execute hers first (insertion is probabilist and not guaranteed). Finally, Mallory could be the \emph{miner} of the block that includes Alice's transaction allowing her to insert with high fidelity. Roll-ups use \emph{sequencers} discussed in Section~\ref{sec:frontarb}.

A \emph{displacement attack} is like an insertion attack, except Mallory does not care what happens to Alice's original transaction---she only cares about being first. If Mallory sees Alice trying to execute a trade at a good price, she could try to beat Alice and execute the trade first. Mallory is indifferent to whether Alice can then execute her trade or not. The analysis of both insertion and suppression attacks are similar.  Call markets mitigate these basic insertion and displacement attacks because they do not have any time priority (e.g., if you were to shuffle the order of all orders submitted within the same call, the outcome would be exactly the same). A different way to mitigate these attacks is to seal orders with confidentiality (a \textit{dark} market).

In a \emph{suppression attack}, Mallory floods the network with transactions until a trader executes her order. Such selective denial of service is possible by an off-chain operator. With on-chain continuous markets, it is not possible to suppress Alice's transaction while also letting through a transaction from a taker---suppression applies to all Ethereum transactions or none. A call market is uniquely vulnerable because it eventually times out (which does not require an on-chain transaction) and new orders cannot be added. We still award a call market partial mitigation since suppression attacks are expensive (\cf Fomo3D attack~\cite{eskandari2019sok}). If the aim of suppression is a temporary denial of service (captured by attack 5 in the table), then all on-chain markets are vulnerable to this expensive attack.

Some attacks combine more than one insertion, displacement, and/or suppression attacks. AMMs are vulnerable to a double insertion called a sandwich attack~\cite{ZQFLG21} which bookends a victim's trade with the front-runner's trades (plus additional variants). In a traditional call market, a market clearing price is chosen and all trades are executed at this price. All bids made at a higher price will receive the assets for the lower clearing price (and conversely for lower ask prices): this is called a \textit{price improvement} and it allows traders to submit at their best price. A hybrid front-running attack allows Mallory to extract any price improvements. Consider the case where Alice's ask crosses Bob's bid with a material price improvement. Mallory inserts a bid at Alice's price, suppresses Bob's bid until the next call, and places an ask at Bob's price. She buys and then immediately sells the asset and nets the price improvement as arbitrage. To mitigate this in \cm, all price improvements are given to the miner (using \texttt{block.coinbase.transfer()}). This does not actively hurt traders---they always receive the same price that they quote in their orders---and it removes any incentive for miners to front-run these profits.

Other front-running attacks use order cancellations (see Section~\ref{sec:cancel}) which \cm mitigates by running short-lived markets with no cancellations.

There are two main takeaways from Table~\ref{tab:front}. Call markets provide strong resilience to front-running only bested slightly by dark markets like TEX~\cite{khalil2019tex}, however, they do it through design---no cryptography and no two-round protocols. A second observation is that dark call markets, like Galal \etal~\cite{galalpublicly}, are no more resilient to front-running than a lit market (however confidentiality could provide resilience to predatory trading algorithms that react quickly to trades without acutally front-running).

\subsection{Front-running on Arbitrum}
\label{sec:frontarb}

In our \cm variant on the Arbitrum, traders can submit transactions to the Layer 1 Inbox contract instead of directly to the \cm DApp. This has the same front-running profile as \cm itself; only the Layer 1 destination address is different. If a sequencer is mandatory, it acts with the same privilege as a Layer 1 Ethereum miner in ordering the transactions it receives. Technically, sequencers are not limited to roll-ups and could be used in the context of normal Layer 1 DApps, but they are more apparent in the context of roll-ups. A sequencer could be trusted to execute transactions in the order it receives them, outsource to a fair ordering service, or (in a tacit acknowledge of the difficulties of preventing front-running) auction off permission to order transactions to the highest bidder (called a \textit{MEV auction}). As shown in Table~\ref{tab:front}, a sequencer is an additional front-running actor but does not otherwise change the kinds of attacks that are possible.




\section{Design Landscape}

\begin{figure}[t]
\centering
\includegraphics[width=1\textwidth]{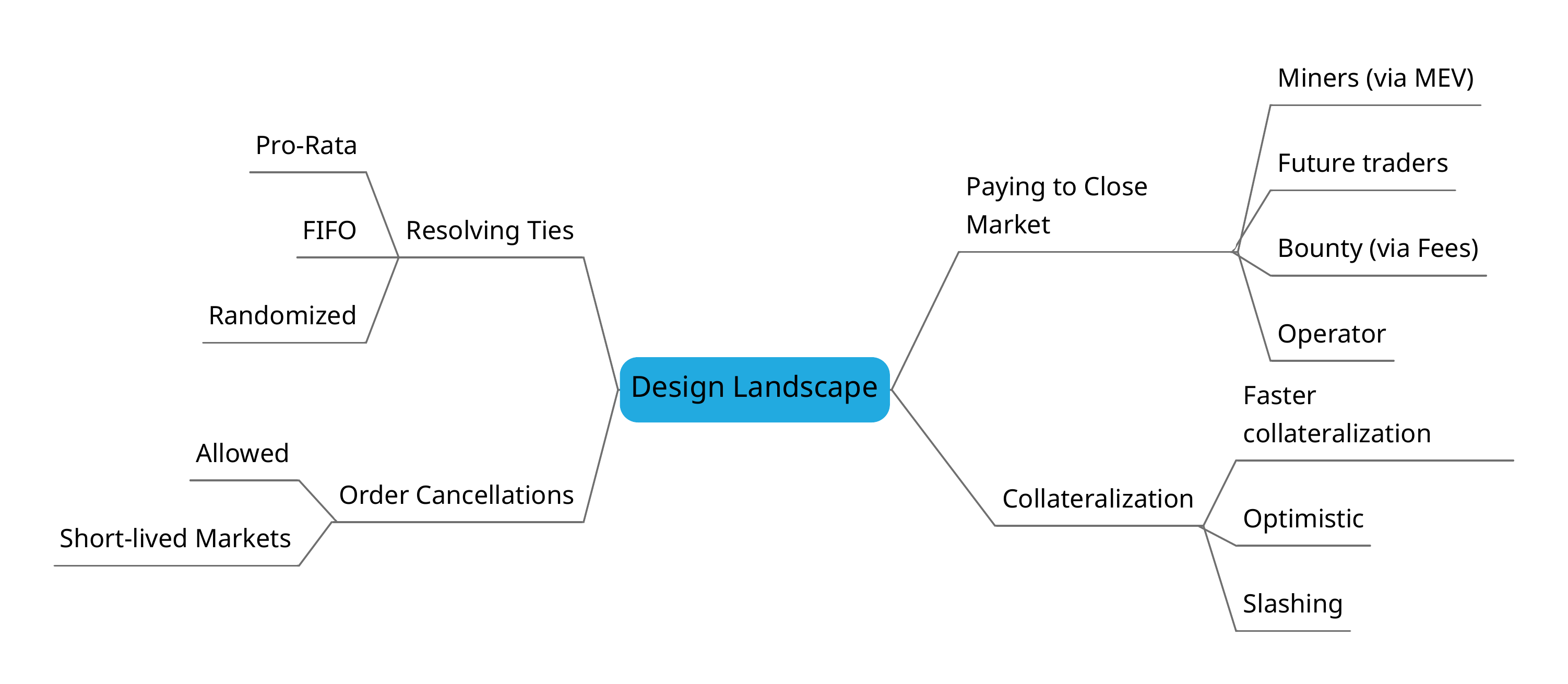}
\caption{\footnotesize{A design landscape for on-chain call markets.}  \label{fig:rmindmap}}
\end{figure}

\cm is a simple base class that implements the core functionality of a call market. To use it in the real world, design decisions need to be made about how it will be used. Figure~\ref{fig:rmindmap} provides a design landscape for \cm deployment, with possible extensions and customization.


\subsection{Token Divisibility and Ties}

A common trading rule is to fill ties in proportion to their volume (\ie \textit{pro rata} allocation)\footnote{If Alice and Bob bid the same price for 100 tokens and 20 tokens respectively, and there are only 60 tokens left in marketable asks, Alice receives 50 and Bob 10.}. This can fail when tokens are not divisible. Consider the following corner case: 3 equally priced bids of 1 non-divisible token and 1 ask at the same price: (1) the bid could be randomly chosen (\cf Libra~\cite{mavroudis2019libra}), or (2) the bid could be prioritized based on time. In \cm, tokens are assumed to be divisible. If the volume of the current best bid does not match the best ask, the larger order is partially filled and the remaining volume is considered against the next best order. We note the conditions under which pro rata allocation fails (\ie non-divisible assets, an exact tie on price, and part of the final allocation) are improbable. (1) is the fairest solution with one main drawback: on-chain sources of `randomness' are generally deterministic and manipulatable by miners~\cite{bonneau2015random,buenz2017proofs}, while countermeasures can take a few blocks to select~\cite{boneh2018verifiable}. We implement (2) which means front-running attacks are possible in this one improbable case. 


\subsection{Order Cancellations}
\label{sec:cancel}

Support for cancellation opens the market to new front-running issues where other traders (or miners) can displace cancellations until after the market closes. However, one benefit of a call market is that beating a cancellation with a new order has no effect, assuming the cancellation is run any time before the market closes. Also, cancellations have a performance impact. Cancelled orders can be removed from the underlying data structure or accumulated in a list that is cross-checked when closing the market. Removing orders requires a more verbose structure than a priority queue (\eg a self-balancing binary search tree instead of a heap; or methods to traverse a linked list rather than only pulling from the head). \cm does not support order cancellations. We intend to open and close markets quickly (on the order of blocks), so orders are relatively short-lived.




\subsection{Who Pays to Close/Reopen the Market?}
\label{sec:close}

In the Princeton paper~\cite{clark2014decentralizing}, the call market is envisioned as an alt-coin, where orders accumulate within a block and a miner closes the market as part of the logic of producing a new block (\ie within the same portion of code as computing their coinbase transaction in Bitcoin or \texttt{gasUsed} in Ethereum). In \cm, someone needs to execute  \texttt{closeMarket()} at the right time and pay for it, which is probably the most significant design challenge for \cm.

Since price improvements are paid to the miners, the miner is incentivized to run \texttt{closeMarket()} if it pays for itself. Efficient algorithms for miners to automatically find `miner extractable value (MEV)' opportunities~\cite{daian2019flash} is an open research problem. Even if someone else pays to close the market, MEV smooths out some market functionality. Assume several orders are submitted and then \texttt{closeMarket()}. A naive miner might order the \texttt{closeMarket()} before the submitted orders, effectively killing those orders and hurting its own potential profit. MEV encourages miners to make sure a profitable \texttt{closeMarket()} in the mempool executes within its current block (to claim the reward for itself) and that it runs after other orders in the mempool to maximize its profit.

Without MEV, markets should open and close on different blocks. In this alternative, the \texttt{closeMarket()} function calls \texttt{openMarket()} as a subroutine and sets two modifiers: orders are only accepted in the block immediately after the current block (\ie the block that executes the \texttt{closeMarket()}) and \texttt{closeMarket()} cannot be run again until two blocks after the current block.

Another option is to have traders in the next call market pay to incrementally close the current market. For example, each order in the next market needs to pay to execute the next $x$ orders in the current market until the order book is empty. This has two issues: first, amortizing the cost of closing the market amongst the early traders of the new market disincentives trading early in the market; the second issue is if not enough traders submit orders in the new market, the old market never closes (resulting in a backlog of old markets waiting to close). 

A closely related option is to levy a carefully computed fee against the traders for every new order they submit. These fees are accumulated by the DApp to use as a bounty. When the time window for the open market elapses, the sender of the first \texttt{closeMarket()} function to be confirmed receives the bounty. This is still not perfect: \texttt{closeMarket()} cost does not follow a tight linear increase with the number of orders, and gas prices vary over time which could render the bounty insufficient for offsetting the \texttt{closeMarket()} cost.  If the DApp can pay for its own functions, an interested party can also arrange for a commercial service (\eg any.sender\footnote{\url{https://github.com/PISAresearch/docs.any.sender}}) to relay the \texttt{closeMarket()} function call on Ethereum (an approach called \textit{meta-transactions}). This creates a regulatory hook.

The final option is to rely on an interested third party (such as the token issuer for a given market) to always close the market, or  occasionally bailout the market when one of the above mechanisms fails. An external service like Ethereum Alarm Clock\footnote{\url{https://ethereum-alarm-clock-service.readthedocs.io/}}  (which also creates a regulatory hook) can be used to schedule regular \texttt{closeMarket()} calls.


\subsection{Collateralization Options}

In \cm, both the tokens and ETH that a trader wants to potentially use in the order book are preloaded into the contract. We discuss some alternative designs in Appendix~\ref{app:collateral}.


\section{Concluding Remarks}

Imagine you have just launched a token on Ethereum. Now you want to be able to trade it. While the barrier to entry for exchange services is low, it still exists. For a centralized or decentralized exchange, you have to convince the operators to list your token and you will be delayed while they process your request. For an automated market maker, you will have to lock up a large amount of ETH into the DApp, along with your tokens. For roll-ups, you will have to host your own servers. By contrast to all of these, with an on-chain order book, you just deploy the code alongside your token and trading is immediately supported. This should concern regulators. Even if it is too slow today, there is little reason for developers not to offer it as a fallback solution that accompanies every token. With future improvements to blockchain scalability, it could become the de facto trading method.

\subsubsection*{Acknowledgements.}  The authors thank the AMF (Autorité des Marchés Financiers) for supporting this research project. J. Clark also acknowledges partial funding from the National Sciences and Engineering Research Council (NSERC)/Raymond Chabot Grant Thornton/Catallaxy Industrial Research Chair in Blockchain Technologies, as well as NSERC through a Discovery Grant. M. Moosavi acknowledges support from Fonds de Recherche du Québec - Nature et Technologies (FRQNT).

\bibliography{bib/DEX}


\appendix

\section{Additional background}
\label{app:back}

\subsection{Ethereum and Blockchain Technology}
\label{app:ethback}

A public blockchain is an open peer-to-peer network that maintains a set of transactions without a single entity in charge. In Ethereum, \emph{transactions} encode the bytecode of user-written \emph{decentralized applications (DApps)} to be stored on the blockchain; and the function calls made to the DApp. Every execution of every function call is validated by all honest, participating nodes to correct; a property that is robust against a fraction of faulty and malicious network nodes (or more precisely, their accumulated computational power). Once transactions are agreed upon, all honest participants will have identical sets of transactions in the same order. For Ethereum, this is conceptualized as the current state of a large \emph{virtual machine (EVM)} that is running many DApps.

Transactions are broadcast by users to the blockchain network where they are propagated to all nodes. Nodes that choose to \emph{mine} will collect transactions (in the order of their choosing) into a block, and will attempt to have the network reach a consensus that their block should be added to the set (or chain) of previous blocks. A transaction is considered finalized once consensus on its inclusion has held for several additional blocks.


\subsubsection{Ethereum's Gas Model.} 

Every transaction results in the participating nodes having to execute bytecode. This is not free. When a transaction is executed, each opcode in the execution path accrues a fixed, pre-specified amount of \emph{gas}. The function caller will pledge to pay a certain amount of Ethereum's internal currency \emph{ETH} (typically quoted in units of Gwei which is one billionth of an ETH) per unit of gas, and miners are free to choose to execute that transaction or ignore it. The function caller is charged for exactly what the transaction costs to execute, and they cap the maximum they are willing to be charged (\textit{gas limit}). If the cap is too low to complete the execution, the miner keeps the Gwei and \emph{reverts} the state of the EVM (as if the function never ran).

A miner can include as many transactions (typically preferring transactions that bid the highest for gas) that can fit under a pre-specified \textit{block gas limit}, which is algorithmically adjusted for every block. As of the time of writing, the limit is approximately 11M gas. Essentially, our main research question is how many on-chain trades can be executed without exceeding that limit. Later, we also discuss several bytecode operations (\emph{opcodes}) that refund gas (\ie cost negative gas), which we heavily utilize in our optimizations.

\subsubsection{Gas Refunds.} 

In order to reconstruct the current state of Ethereum's EVM, a node must obtain a copy of every variable change since the genesis block (or a more recent `checkpoint' that is universally agreed to). For this reason, stored variables persist for a long time and, at first glance, it seems pointless to free up variable storage (and unclear what `free up' even means). Once the current state of the EVM is established by a node, it can forget about every historical variable changes and only concern itself with the variables that have non-zero value (as a byte string for non-integers) in the current state (uninitialized variables in Ethereum have the value 0 by default). Therefore, freeing up variables will reduce the amount of state Ethereum nodes need to maintain going forward.

For this reason, some EVM operations cost a negative amount of gas. That is, the gas is \textit{refunded} to the sender at the end of the transaction, however (1) the refund is capped at 50\% of the total gas cost of the transaction, and (2) the block gas limit applies to the pre-refunded amount (\ie a transaction receiving a full refund can cost up to 5.5M gas with an 11M limit). Negative gas operations include:

\begin{itemize}

\item \texttt{SELFDESTRUCT}. This operation destroys the contract that calls it and refunds its balance (if any) to a designated receiver address. The  \texttt{SELFDESTRUCT} operation does not remove the initial byte code of the contract from the chain. It always refunds 24,000 gas. For example, if contract A stores a single non-zero integer and contract B stores 100 non-zero integers, the \texttt{SELFDESTRUCT} refund for both is the same (24,000 gas).

\item \texttt{SSTORE}. This operation loads a storage slot with a value. Using \texttt{SSTORE} to load a zero into a storage slot with a non-zero value means the nodes can start ignoring it (recall that all variables, even if uninitialized, have zero by default). Doing this refunds 15,000 gas per slot.

\end{itemize}

At the time of this writing, Ethereum transaction receipts only account for the \texttt{gasUsed}, which is the total amount of gas units spent during a transaction, and users are not able to obtain the value of the EVM's refund counter from inside the EVM~\cite{signer2018gas}. So in order to account for refunds in Table~\ref{tab:PQUnitTests}, we calculate them manually. First, we determine exactly how many storage slots are being cleared or how many smart contracts are being destroyed, then we multiply these numbers by 24,000 or 15,000 respectively.

\subsubsection{Optimistic Roll-Ups.}
\label{app:rollup}

\begin{figure}[htb]
  \centering
  \begin{turn}{90}
  \begin{minipage}{8in}
  \centering
\includegraphics[width=1\textwidth]{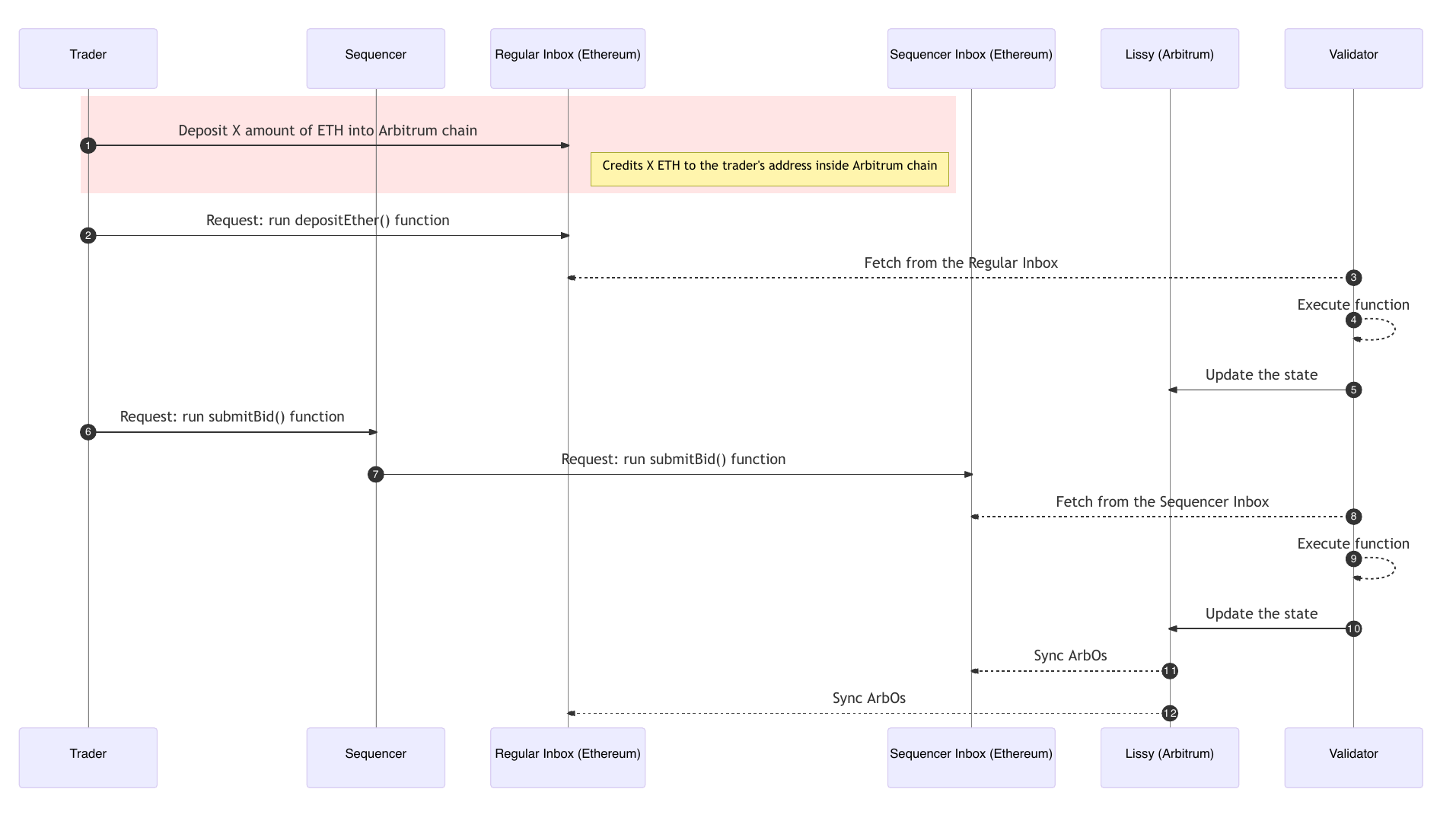}
  \caption{\footnotesize{Overview of \cm on Arbitrum.}
\label{fig:lissyl2}}
  \end{minipage}
  \end{turn}
\end{figure}


\textit{Layer 2} solutions are a group of technologies that are designed and proposed to address specific drawbacks of executing transactions on \textit{Layer 1} (\ie Ethereum and other blockchains)~\cite{gudgeon2020sok}. These technologies focus on fast transaction throughput, reducing gas costs, or educing transaction latency. When using \cm, we strive to reduce the gas cost as performance is the main bottleneck. Thus, we choose a Layer 2 technology called \textit{roll-up} which aims at reducing the gas cost for operating on Layer 1 by taking the transaction executions off-chain and only using the Ethereum blockchain for storing data. In a roll-up, every transaction is executed by a server or cluster of servers known as \textit{validators} that can be run by a collection of users or third party operators (here they can be run by the token issuer). These validators then push the result of the executions (\ie updates in the EVM state) back to the Ethereum and assure the Ethereum network that the transactions have been executed correctly.

A function can be computed off-chain and the new state of the DApp, called a \textit{rollup}, is written back to the blockchain, accompanied by either (1) a proof that the function was executed correctly, or (2) a dispute resolution process that can resolve, on-chain, functions that are not executed correctly (\eg Arbitrum~\cite{kalodner2018arbitrum}). In the case of (1), validating the proof must be cheaper than running the function itself. There are two main approaches: (1a) the first is to use cryptographic proof techniques (\eg SNARKS~\cite{BCGTV13,GGPR13} and variants~\cite{BBHR19}). This is called a \textit{zk-rollup}. Note that the proofs are heavy to compute (introducing a burden to the validators who generate them) but considered valid once posted to the Ethereum. The second approach (1b) is to execute the function in a trusted execution environment (TEE; \eg Intel SGX) and validate the TEE's quote on-chain (\eg Ekiden~\cite{cheng2019ekiden}).\footnote{The TEE-based approach is mired by recent attacks on SGX~\cite{SGX1,SGX2,SGX3,SGX4}, however these attacks do not necessarily apply to the specifics of how SGX is used here, and safer TEE technologies like Intel TXT (\cf~\cite{ZBC+19}) can be substituted.} Approach (2) is called an \textit{optimistic roll-up}. Although the dispute time delays result in a slower transaction finality, optimistic roll-ups substantially increase the performance by decreasing the gas cost. 

Arbitrum and Ethereum Optimism are the two prominent deployments of an optimistic roll-up. Arbitrum  uses a multi-round dispute process that results in very minimal L1 gas costs to resolve a dispute. Specifically, if a dispute over a transaction arrises, the L1 cost of resolving the dispute is a small fraction of the cost of executing the transaction itself (whereas in Optimism, the dispute resolution cost is essentially the same as executing the transaction). 

Figure~\ref{fig:lissyl2} shows how traders interact with \cm on Arbitrum. First, a trader sends a \texttt{depositETH} transaction on Ethereum to the Inbox contract to deposit \textit{X} amount of ETH to the Arbitrum chain. Once the transaction is confirmed, \textit{X} amount of ETH will be credited to the trader's address on the Arbitrum chain. Trader can now interact with \cm and execute its functions by sending the instruction and data required for those executions to either (1) the Arbitrum regular Inbox on Ethereum, or (2) the sequencer. In our example, trader uses the regular Inbox to execute \texttt{depositEther()} and the sequencer to execute \texttt{submitBid()} from \cm that lives entirely on Arbitrum chain. 
Accordingly, trader deposits ETH to \cm smart contract by sending the instruction and data for executing the \texttt{depositEther()} to the Arbitrum Inbox contract that lives on Ethereum. A validator fetches this transaction from the Inbox, executes it, and asserts the result to ArbOS. Next, trader sends the instruction and data for execution of \texttt{submitBid()} to the sequencer. The sequencer then inserts this message into the Inbox that it owns. This Inbox contract has the same interface as the regular Inbox contract, however, it is \textit{owned} by the sequencer. A validator sees the transaction in the sequencer Inbox of the bridge, executes it,  and asserts the result to ArbOS. Periodically, the entire state of ArbOS is committed back to Ethereum. 

Our \cm variant is not the first roll-up-based order book. Loopring 3.0\footnote{\url{https://loopring.org}} offers a continuous-time order book. The primary difference is that orders in Loopring 3.0 are submitted off-chain to the operator directly, whereas our variant uses on-chain submission so that the roll-up server does not need to be publicly reachable. Loopring 3.0 can operate near high-frequency trading as order submission is unhampered by Ethereum. However, its  roll-up proof does not ensure that the exchange did not reorder transactions, which is particularly problematic in a continuous-time order book. Traders who prioritize trade fairness might opt for a solution like our variant, while traders who want speed would vastly prefer the Loopring architecture which offers near-CEX speed while being non-custodial. Loopring leaves a regulatory hook whereas our variant could be nearly as difficult to regulate as a fully on-chain solution if the roll-up server was kept anonymous: Ethereum and Arbitrum themselves would be the only regulatory hooks.

\subsection{Trade Execution Systems}
\label{app:markets}

\paragraph{Centralized Exchanges (CEX).} Traditional financial markets (\eg NYSE and NASDAQ) use order-matching systems to arrange trades. An exchange will list one or more assets (stocks, bonds, derivatives, or more exotic securities) to be traded with each other, given its own order book priced in a currency (\eg USD). Exchanges for blockchain-based assets (also called crypto assets by enthusiasts) can operate the same way, using a centralized exchange (CEX) design where a firm (\eg Binance, Bitfinex, \etc) operates the platform as a trusted third party in every aspect: custodianship over assets/currency being traded, exchanging assets fairly, offering the best possible price execution. Security breaches and fraud in centralized exchanges  (\eg MtGox~\cite{TheHisto45:online}, QuadrigaCX~\cite{SEBIOrde83:online}, and many others) have become a common source of lost funds for users, while accusations of unfair trade execution have been leveled but are difficult to prove. Today, CEXes are often regulated as other money service businesses---this provides some ability for the government to conduct financial tracking but does little to provide consumer protection against fraud.

\paragraph{On-chain Order Books.} For trades between two blockchain-based assets (\eg a digital asset priced in a cryptocurrency, stablecoin, or second digital asset), order matching can be performed `on-chain' by deploying the order-matching system either on a dedicated blockchain or inside a decentralized application (DApp). In this model, traders entrust their assets to an autonomously operating DApp with known source code instead of a third party custodian that can abscond with or lose the funds. The trading rules will operate as coded, clearing and settling can be guaranteed, and order submission is handled by the blockchain---a reasonably fair and transparent system (but see front-running below). Finally, anyone can create an on-chain order book for any asset (on the same chain) at any time. While these sound ideal, performance is a substantial issue and the main subject of this paper. Since it is an open system, there is no obvious regulatory hook (beyond the blockchain itself).

In this paper, we focus on benchmarking an order book for the public blockchain Ethereum. Ethereum is widely used and we stand to learn the most from working in a performance-hostile environment. Exchanges could be given their own dedicated blockchain, where trade execution logic can be coded into the network protocol. Trading systems on permissioned blockchains (\eg NASDAQ Linq, tZero) can also improve execution time and throughput, but they reduce user transparency and trust if unregulated.

\paragraph{On-chain Dealers.} An advantage of on-chain trading is that other smart contracts, not just human users, can initiate trades, enabling broader decentralized finance (DeFi) applications. This has fueled a resurgence in on-chain exchange but through a quote-driven design rather than an order-driven one. Automated market makers  (\eg Uniswap v3) have all the trust advantages of an on-chain order book, plus they are relatively more efficient. The trade-off is that they operate as a dealer---the DApp exchanges assets from its own inventory. This inventory is loaded into the DApp by an investor who will not profit from the trades themselves but hopes their losses (termed `impermanent losses') are offset over the long-term by trading fees. By contrast, an order book requires no upfront inventory and trading fees are optional. Finally, there is a complicated difference in their price dynamics (\eg market impact of a trade, slippage between the best bid/ask and actual average execution price, \etc)---deserving of an entire research paper to precisely define. We leave it as an assertion that with equal liquidity, order books have more favorable price dynamics for traders.

\paragraph{Hybrid Designs.} Before on-chain dealers became prominent in the late 2010s, the most popular design was hybrid order-driven exchanges with some trusted off-chain components and some on-chain functionalities. Such decentralized exchanges (DEXes) were envisioned as operating fully on-chain, but performance limitations drove developers to move key components, such as the order matching system, off-chain to a centralized database. A landscape of DEX designs exist (\eg EtherDelta, 0x, IDEX, \etc): many avoid taking custodianship of assets off-chain, and virtually all (for order-driven markets) operate the order book itself off-chain (a regulatory hook). A non-custodial DEX solves the big issue of a CEX---the operator stealing the funds---however trade execution is still not provably fair, funds can still be indirectly stolen by a malicious exchange executing unauthorized trades, and server downtime is a common frustration for traders. An enhancement is to prove that trade execution is correct (\eg Loopring) but these proofs have blind spots (discussed above in Appendix~\ref{app:rollup}).

\subsection{Call Markets}
Assume traders submit their orders in Table~\ref{tab:cmsample} to a call market when it is open. In the following, we explain how these orders are executed:

\begin{table}[t]
\centering
\scriptsize
\begin{tabular}{|c|c|c|c|c|c}
\hline

\textbf{Time}& \textbf{Trader}   & \textbf{Order Type} & \textbf{Order Price}   & \textbf{Volume}     \\ \hline

09:10   & Mehdi & Ask & 10.18 &4              \\ \hline
09:12   & Avni & Bid & 12 &3              \\ \hline
09:15   & Kritee & Bid & 13 &3              \\ \hline
09:18   & Bob & Bid & 12.15 &1              \\ \hline
09:26   & Navjot & Ask & 10.15 &4              \\ \hline
09:30   & Alice & Ask & 10 &1              \\ \hline

\end{tabular}
\caption{\footnotesize{Example orders that are submitted to a call market}.
\label{tab:cmsample}}
\end{table}

\label{app:cm}
\begin{itemize}
\item The call market first matches Alice's ask order to sell 1 at 10 with Avni's bid order to buy 3 at 12. Trade occurs at the price Alice asks for; 10, and 2 will be given to the miner as a price improvement. This trade fills Alice's order and leaves Avni with a remainder of 2 to buy at 12. 
\item Next, the call market matches Avni's remainder of 2 with the next highest priority ask order in the list which is Navjot's order to sell 4 at 10.15. Trade occurs at 10.15 and 1.85 will be given to the miner as a price improvement. This trade fills the remainder of Avni's bid order and leaves Navjot with a remainder of 2 to sell at 10.15.
\item The market now matches the next highest bid order in the list, Bob's bid order to buy 1 at 12.15, with the remainder of Navjot's ask order to sell 2 at 10.15. Trader occurs at 10.15 and 2 will be given to miner as a price improvement. This trade fills Bob's bid order and leaves Navjot with a remainder of 1 to sell at 10.15.
\item Next, the market matches Kritee's bid order to buy 3 at 13 with the remainder of Navjot's ask order to sell 1 at 10.15. Trade occurs at 10.15 and 2.85 will be given to miner as a price improvement. This trade fills Navjot's order and leaves Kritee with a remainder of 2 to buy at 13.
\item The market then matches Mehdi's ask order to sell 4 at 10.18 with the remainder of Kritee's bid order to buy 2 at 13. Trade occurs at 10.18 and 2.82 is given to miner as a price improvement. This trade fills Kritee's order and leaves Mehdi with a remainder of 2 to sell at 10.18 unfilled. 
\end{itemize}

\section{Cleaning-Up Revisited: Clearing Mappings}
\label{app:clean}

Beyond the cleaning up issues with priority queues in Section~\ref{sec:gasrefund}, \cm also uses mappings with each market. Traders preload their account with tokens to be traded (which comply with a common token standard called ERC20) and/or ETH. \cm tracks what they are owed using a mapping called \texttt{totalBalance} and allows traders to withdraw their tokens at any time. However if a trader submits an order (\ie ask for their tokens), the tokens are committed and not available for withdrawal until the market closes (after which, the balances are updated for each trade that is executed). Committed tokens are also tracked in a mapping called \texttt{unavailableBalance}. Sellers can request a token withdrawal up to their total balance subtracted by their unavailable balance.

As the DApp runs \texttt{closeMarket()}, it starts matching the best bids to the best asks. As orders execute, \texttt{totalBalance} and \texttt{unavailableBalance} are updated. At a certain point, the bids and asks will stop matching in price. At this point, every order left in the order book cannot execute (because the priority queue sorts orders by price, and so orders deeper in the queue have worst prices than the order at the head of the queue). Therefore all remaining entries in \texttt{unavailableBalance} can be cleared.

In Solidity, it is not possible to delete an entire mapping without individually zeroing out each entry key-by-key. At the same time, it is wasteful to let an entire mapping sit in the EVM when it will never be referenced again. The following are some options for addressing this conflict.

\begin{enumerate}

\item \textbf{Manually Clearing the Mapping.} Since mappings cannot be iterated, a common design pattern used by DApp developers is to store keys in an array and iterate over the array to zero out each mapping and array entry. Clearing a mapping this way costs substantially more to clear than what is refunded.

\item \textbf{Store the Mapping in a Separate DApp.} We could wrap the mapping inside its own DApp and when we are done with the mapping, we can run \texttt{SELFDESTRUCT} on the contract. This refunds us 24,000 gas which is less than the cost of deploying the extra contract. Additionally, every call to the mapping is more expensive because (1) it is an external function call, and (2) the calls need access control to ensure only the market contract can write to it (if a mapping is a local variable, you get private access for free). 

\item \textbf{Leave and Ignore the Mapping.} The final option is to not clear the mapping and just create a new one (or create a new prefix for all mapping keys to reflect the new version of the mapping). Unfortunately, this is the most economical option for DApp developers even if it is the worst option for Ethereum nodes. 

\end{enumerate}

Clearing storage is important for reducing EVM bloat. The Ethereum refund model should be considered further by Ethereum developers to better incentivize developers to be less wasteful in using storage. 


\section{Collateralization Options in Call Markets}
\label{app:collateral}

in \cm, both the tokens and ETH that a trader wants to potentially use in the order book are preloaded into the contract. Consider Alice, who holds a token and decides she wants to trade it for ETH. In this model, she must first transfer the tokens to the contract and then submit an ask order. If she does this within the same block, there is a chance that a miner will execute the ask before the transfer and the ask will revert. If she waits for confirmation, this introduces a delay. This delay seems reasonable but we point out a few options it could be addressed:

\begin{enumerate}

\item \textbf{Use \texttt{msg.value}.} For the ETH side of a trade (\ie for bids), ETH could be sent with the function call to \texttt{submitBid()} to remove the need for    \texttt{depositEther()}. This works for markets that trade ERC20 tokens for ETH, but would not work for ERC20 to ERC20 exchanges.

\item \textbf{Merge Deposits with Bids/Asks.} \cm could have an additional function that atomically runs the functionality of \texttt{depositToken()} followed by the functionality of \texttt{submitAsk()}. This removes the chance that the deposit and order submission are ordered incorrectly.

\item \textbf{Use ERC20 Approval.} Instead of \cm taking custody of the tokens, the token holder could simply approve \cm to transfer tokens on her behalf. If \cm is coded securely, it is unconcerning to allow the approval to stand long-term and the trader never has to lock up their tokens in the DApp. The issue is that there is no guarantee that the tokens are actually available when the market closes (\ie Alice can approve a DApp to spend 100 tokens even if she only has 5 tokens or no tokens). In this case, \cm would optimistically try to transfer the tokens and if it fails, move onto the next order. This also gives Alice an indirect way to cancel an order, by removing the tokens backing the order---this could be a feature or it could be considered an abuse.

\item \textbf{Use a Fidelity Bond.} Traders could post some number of tokens as a fidelity bond, and be allowed to submit orders up to 100x this value using approve. If a trade fails because the pledged tokens are not available, the fidelity bond is slashed as punishment. This allows traders to side-step time-consuming transfers to and from \cm while still incentivizing them to ensure that submitted orders can actually be executed. The trade-off  is that \cm needs to update balances with external calls to the ERC20 contract instead of simply updating its internal ledger.

\end{enumerate}

\section{Market Clearing Prices}


Call markets are heralded for fair price discovery. This is why many exchanges use a call market at the end of the day to determine the closing price of an asset, which is an important price both optically (it is well published) and operationally (many derivatives settle based on the closing price). We purposely do not compute a `market clearing price' with \cm because miners can easily manipulate the price (\ie include a single wash trade at the price they want fixed), although they forgo profit for doing so. This is not merely hypothetical---Uniswap (the prominent quote-drive, on-chain exchange) prices have been manipulated to exploit other DeFi applications relying on them. Countermeasures to protect Uniswap price integrity could also apply to \cm: (1) taking a rolling median of prices over time, and (2) using it alongside other sources for the same price and forming a consensus. While \cm does not emit a market clearing price, it can be computed by a web application examining the order book at market close.

\end{document}